\documentclass{ws-procs9x6}

\newlength{\dslashwidth}

\newcommand{\bq}{\begin{equation}}
\newcommand{\eq}{\end{equation}}
\newcommand{\ba}{\begin{array}}
\newcommand{\ea}{\end{array}}
\newcommand{\bqa}{\begin{eqnarray}}
\newcommand{\eqa}{\end{eqnarray}}

\newcommand{\lnf}{{\ifmmode \Lambda^{(N_f)} \else $\Lambda^{(N_f)}$\fi}}
\newcommand{\ms}{{\ifmmode \overline{MS} \else $\overline{MS}$\fi}}
\newcommand{\dr}{{\ifmmode \overline{DR} \else $\overline{DR}$\fi}}
\newcommand{\lms}{{\ifmmode \Lambda^{(5)}_{\overline{MS}} \else $\Lambda^{(5)}_{\overline{MS}}$\fi}}
\newcommand{\lam}{{\ifmmode \Lambda \else $\Lambda$\fi}}
\newcommand{\gev}{{\ifmmode {\rm GeV} \else ${\rm GeV}$\fi}}
\newcommand{\gevc}{{\ifmmode {\rm GeV/c^2} \else ${\rm GeV/c^2}$\fi}}
\newcommand{\tev}{{\ifmmode {\rm TeV} \else ${\rm TeV}$\fi}}
\newcommand{\tevc}{{\ifmmode {\rm TeV/c^2} \else ${\rm TeV/c^2}$\fi}}
\newcommand{\lp}{{\ifmmode L^+  \else $L^+$\fi}}
\newcommand{\lm}{{\ifmmode L^-  \else $L^-$\fi}}
\newcommand{\mlp}{{\ifmmode M(L^-) \else $M(L^-)$\fi}}
\newcommand{\mlz}{{\ifmmode M(L^0) \else $M(L^0)$\fi}}
\newcommand{\lz}{{\ifmmode L^0 \else $L^0$\fi}}
\newcommand{\ev}{{\ifmmode GeV/c^2 \else $GeV/c^2$\fi}}
\newcommand{\tri}{{\ifmmode \triangleup \else $\triangleup$\fi}}
\newcommand{\unl}{{\ifmmode U_{lL^0} \else $U_{lL^0}$\fi}}\newcommand{\gL}{{\ifmmode g_L \else $g_{L}$\fi}}
\newcommand{\gR}{{\ifmmode g_R  \else $g_{R}$\fi}}
\newcommand{\gumu}{{\ifmmode \gamma^{\mu} \else $\gamma^{\mu}$\fi}}
\newcommand{\gunu}{{\ifmmode \gamma^{\nu} \else $\gamma^{\nu}$\fi}}
\newcommand{\gdmu}{{\ifmmode \gamma_{\mu} \else $\gamma_{\mu}$\fi}}
\newcommand{\gdnu}{{\ifmmode \gamma_{\nu} \else $\gamma_{\nu}$\fi}}
\newcommand{\stw}{{\ifmmode\sin^2\theta_W \else $\sin^{2}\theta_{W}$ \fi}}
\newcommand{\sws}{{\ifmmode \;\sin^2\theta_W  \else $\;\sin^{2}\theta_{W}$ \fi}}
\newcommand{\cws}{{\ifmmode \;\cos^2\theta_W  \else $\;\cos^{2}\theta_{W}$ \fi}}
\newcommand{\sw}{{\ifmmode \;\sin\theta_W  \else $\sin\theta_{W}$ \fi}}
\newcommand{\cw}{{\ifmmode \;\cos\theta_W  \else $\;\cos\theta_{W}$ \fi}}
\newcommand{\tw}{{\ifmmode \;\tan\theta_W  \else $\;\tan\theta_{W}$ \fi}}
\newcommand{\qq}{{\ifmmode q\overline{q} \else $q\overline{q}$\fi}}
\newcommand{\lR}{{\ifmmode l_R \else $l_R$\fi}}
\newcommand{\lL}{{\ifmmode l_L \else $l_L$\fi}}
\newcommand{\nt}{{\ifmmode \nu_{\tau} \else $\nu_{\tau}$\fi}}
\newcommand{\nuR}{{\ifmmode \nu_R  \else $\nu_R$\fi}}
\newcommand{\nuL}{{\ifmmode \nu_L  \else $\nu_L$\fi}}
\newcommand{\qR}{{\ifmmode g_R  \else $q_R$\fi}}
\newcommand{\qL}{{\ifmmode q_L  \else $q_L$\fi}}
\newcommand{\qRp}{{\ifmmode q_R'  \else $q_{R}$'\fi}}
\newcommand{\qLp}{{\ifmmode q_L'  \else $q_{L}$'\fi}}
\newcommand{\est}{{\ifmmode e^{\bf \ast} \else $e^{\bf \ast}$\fi}}
\newcommand{\lst}{{\ifmmode l^{\bf \ast} \else $l^{\bf \ast}$\fi}}
\newcommand{\must}{{\ifmmode \mu^{\bf \ast} \else $\mu^{\bf \ast}$\fi}}
\newcommand{\taust}{{\ifmmode \tau^{\bf \ast} \else $\tau^{\bf \ast}$ \fi}}
\newcommand{\pperp}{{\ifmmode p_t  \else $p_t$\fi}}
\newcommand{\et}{{\ifmmode E_t  \else $E_t$\fi}}
\newcommand{\xt}{{\ifmmode x_t  \else $x_t$\fi}}
\newcommand{\smumu}{{\ifmmode \sigma_{\mu\mu}  \else $\sigma_{\mu\mu}$ \fi}}
\newcommand{\eg}{{\ifmmode e\gamma  \else $e\gamma$\fi}}
\newcommand{\epem}{{\ifmmode e^+e^-  \else $e^+e^-$\fi}}
\newcommand{\lplm}{{\ifmmode L^+L^-  \else $L^+L^-$\fi}}
\newcommand{\pp}{{\ifmmode p\overline p  \else $p\overline p$\fi}}
\newcommand{\llz}{{\ifmmode L^0\overline{L}^0 \else $L^0\overline{L}^0$\fi}}
\newcommand{\epemt}{{\ifmmode e^+e^- \to  \else $e^+e^- \to$\fi}}
\newcommand{\eb}{{\ifmmode E_{beam}  \else $E_{beam}$\fi}}
\newcommand{\ip}{{\ifmmode pb^{-1}  \else $pb^{-1}$\fi}}
\newcommand{\upm}{{\ifmmode ^{\pm}  \else $^{\pm}$\fi}}
\newcommand{\de}{{\ifmmode ^{\circ}  \else $^{\circ}$ \fi}}
\newcommand{\appr}{{\ifmmode \sim \else $\sim$ \fi}}
\newcommand{\corresp}{{\ifmmode \stackrel{\wedge}{=} \else $\stackrel{\wedge}{=}$ \fi}}
\newcommand{\sqrts}{{\ifmmode \sqrt{s} \else $\sqrt{s}$\fi}}
\newcommand{\zz}{{\ifmmode Z^0  \else $Z^0$\fi}}
\newcommand{\mz}{{\ifmmode M_{Z}  \else $M_{Z}$\fi}}
\newcommand{\mzs}{{\ifmmode M_{Z}^2  \else $M_{Z}^2$\fi}}
\newcommand{\mw}{{\ifmmode M_{W}  \else $M_{W}$\fi}}
\newcommand{\mws}{{\ifmmode M_{W}^2  \else $M_{W}^2$\fi}}
\newcommand{\mh}{{\ifmmode M_{Higgs}  \else $M_{Higgs}$\fi}}
\newcommand{\gt}{{\ifmmode \Gamma_{tot} \else $\Gamma_{tot}$\fi}}
\newcommand{\msusy}{{\ifmmode M_{SUSY}  \else $M_{SUSY}$\fi}}
\newcommand{\msusys}{{\ifmmode M_{SUSY}^2  \else $M_{SUSY}^2$\fi}}
\newcommand{\su}{{\ifmmode SU(3)_C\otimes\- SU(2)_L\otimes\- U(1)_Y \else $SU(3)_C\otimes SU(2)_L\otimes U(1)_Y$\fi}}
\newcommand{\suthree}{{\ifmmode SU(3)_C  \else $SU(3)_C$\fi}}
\newcommand{\sutwo}{{\ifmmode  SU(2)_L\otimes U(1)_Y \else $SU(2)_L\otimes U(1)_Y$\fi}}
\newcommand{\taup} {{\ifmmode \tau_{proton} \else $\tau_{proton}$\fi}}
\newcommand{\as}{{\ifmmode \alpha_{s}  \else $\alpha_{s}$\fi}}
\newcommand{\mgut}{{\ifmmode M_{GUT}  \else $M_{GUT}$\fi}}
\newcommand{\mguts}{{\ifmmode M_{GUT}^2  \else $M_{GUT}^2$\fi}}
\newcommand{\mze} {{\ifmmode m_0        \else $m_0$\fi}}
\newcommand{\mha}{{\ifmmode m_{1/2}    \else $m_{1/2}$\fi}}
\newcommand{\mb} {{\ifmmode m_{b}    \else $m_{b}$\fi}}
\newcommand{\mt} {{\ifmmode m_{t}    \else $m_{t}$\fi}}
\newcommand{\mts} {{\ifmmode m_{t}^2    \else $m_{t}^2$\fi}}

\newcommand{\mtau}{{\ifmmode m_{\tau}  \else $m_{\tau}$\fi}}
\newcommand{\dpp}{{\ifmmode \delta_{pert} \else $\delta_{pert}$\fi}}
\newcommand{\dnp}{{\ifmmode\delta_{non-pert}\else$\delta_{non-pert}$\fi}}
\newcommand{\dew}{{\ifmmode \delta_{\rm EW}\else $\delta_{\rm EW}$\fi}}
\newcommand{\rt}{{\ifmmode R_{\tau}  \else $R_{\tau} $\fi}}
\newcommand{\rz}{{\ifmmode R_{Z}  \else $R_{Z} $\fi}}

\newcommand{\swb}{{\ifmmode \sin^2\theta_{\overline{MS}} \else $\sin^2\theta_{\overline{MS}}$\fi}}
\newcommand{\cwb}{{\ifmmode \cos^2\theta_{\overline{MS}} \else $\cos^2\theta_{\overline{MS}}$\fi}}

\begin{document}

\title{Is the Dark Matter interpretation of the EGRET gamma ray excess
compatible with antiproton measurements?}

\author{W. de Boer, I. Gebauer, C. Sander, M. Weber, V. Zhukov}

\address{Institut f\"ur Experimentelle Kernphysik\\
University of Karlsruhe,
Physikhochhaus\\
Postfach 6980,
76128 Karlsruhe, Germany\\
E-mail: wim.de.boer@cern.ch}

\begin{abstract}
The diffuse galactic EGRET gamma ray data
 show a clear excess for energies above 1 GeV in comparison
with the expectations from conventional galactic models. This excess
shows all the features expected from Dark Matter WIMP  Annihilation:
a)it is present and has the same spectrum in all sky directions, not
just in the galactic plane, as expected for WIMP  annihilation b) it
shows an interesting substructure  in the form of a doughnut shaped
ring at 14 kpc from the centre of the galaxy, where a ring of stars
indicated the probable infall of a dwarf galaxy. From the spectral
shape of the excess the WIMP mass is estimated to be between 50 and
100 GeV, while from the intensity the halo profile is reconstructed,
which is shown to explain the peculiar change of slope in the
rotation curve at about 11 kpc (due to the ring of DM at 14 kpc).

Recently it was claimed by Bergstr$\rm\ddot{o}$m et al. that the DM
interpretation of the EGRET gamma ray excess is excluded by the
antiproton fluxes, since in their propagation model with isotropic
diffusion the flux of antiprotons would be far beyond the observed
flux. However, the propagation can be largely anisotropic, because
of the convection of particles perpendicular to the disc and
inhomogeneities in the local environment. It is shown that
anisotropic propagation can reduce the antiproton yield by an order
of magnitude, while still being consistent with the B/C ratio.
 Therefore it is hard to use antiprotons to  search for
{\it light} DM particles, which yield a similar antiproton spectrum
as the background, but the antiprotons are a perfect means  to tune
the many degenerate parameters in the propagation models.
\end{abstract}
\bodymatter
\section{Introduction}
Cold Dark Matter (CDM) makes up 23\% of the energy of the universe,
as deduced from the WMAP measurements of the temperature
anisotropies in the Cosmic Microwave Background, in combination with
data on the Hubble expansion and the density fluctuations in the
universe~\cite{wmap}. The nature of the CDM is unknown, but one of
the most popular explanation for it is the neutralino, a stable
neutral particle predicted by Supersymmetry~\cite{jungman}. The
neutralinos are spin 1/2 Majorana particles, which can annihilate
into pairs of Standard Model (SM) particles. The stable decay and
fragmentation products are neutrinos, photons, protons, antiprotons,
electrons and positrons. From these, the protons and electrons
disappear in the sea of many matter particles in the universe, but
the photons and antimatter particles may be detectable above the
background, generated by  particle interactions. Searches for the
stable products of dark matter annihilation (DMA) (so-called
indirect Dark Matter detection) have been actively pursued, see e.g
the review by Bergstr$\rm\ddot{o}$m\cite{bergstrom} or more recently
by Bertone, Hooper and Silk \cite{hooper}.

In previous papers  we showed that the so-called EGRET excess of
diffuse galactic gamma rays\cite{hunter} exhibits all the features
of DMA.\cite{deboer1,deboer2,deboer3} However,
Bergstr$\rm\ddot{o}$m et al. \cite{bergstrom_pb} claimed that the DM
interpretation of the EGRET gamma ray excess is incompatible with
the antiproton fluxes, since in their propagation model  with
isotropic diffusion (based on DarkSusy) the flux of antiprotons
would be far beyond the observed flux. In this contribution it is
shown that more realistic propagation models  could solve this
problem.

After summarizing the DMA interpretation of the excess of gamma
rays, the expected antiproton flux will be discussed based on the
GALPROP propagation model\cite{galprop} after implementing and
retuning its parameters and taking into account the expected
anisotropic propagation and the clumpiness of the gas distribution.

\section{Gamma rays from Dark Matter Annihilation}
The thermally averaged
 annihilation cross section for any thermal relic is known from the
 inverse proportionality to the relic density~\cite{jungman}.
 This cross section comes out to be that of a Weakly
 Interacting Massive Particle (WIMP). The dark matter annihilation (DMA)
   is expected to yield
 predominantly  mono-energetic quark pairs, since the kinetic energy
 is negligible for CDM.
From the hadronization of the quarks one expects a large flux of
gamma rays from the decay of the $\pi_0$ mesons, typically several
tens of gamma rays per annihilation with energies of several GeV.
The gamma ray spectrum from mono-energetic quarks has been studied
in detail in the hadronization of quarks produced at
electron-positron colliders. The DMA gamma ray spectrum is
considerably harder than the background spectrum, which originates
from inelastic scattering of cosmic rays (CR) on the interstellar
gas.
\begin{figure}[t]
\includegraphics[width=0.48\textwidth]{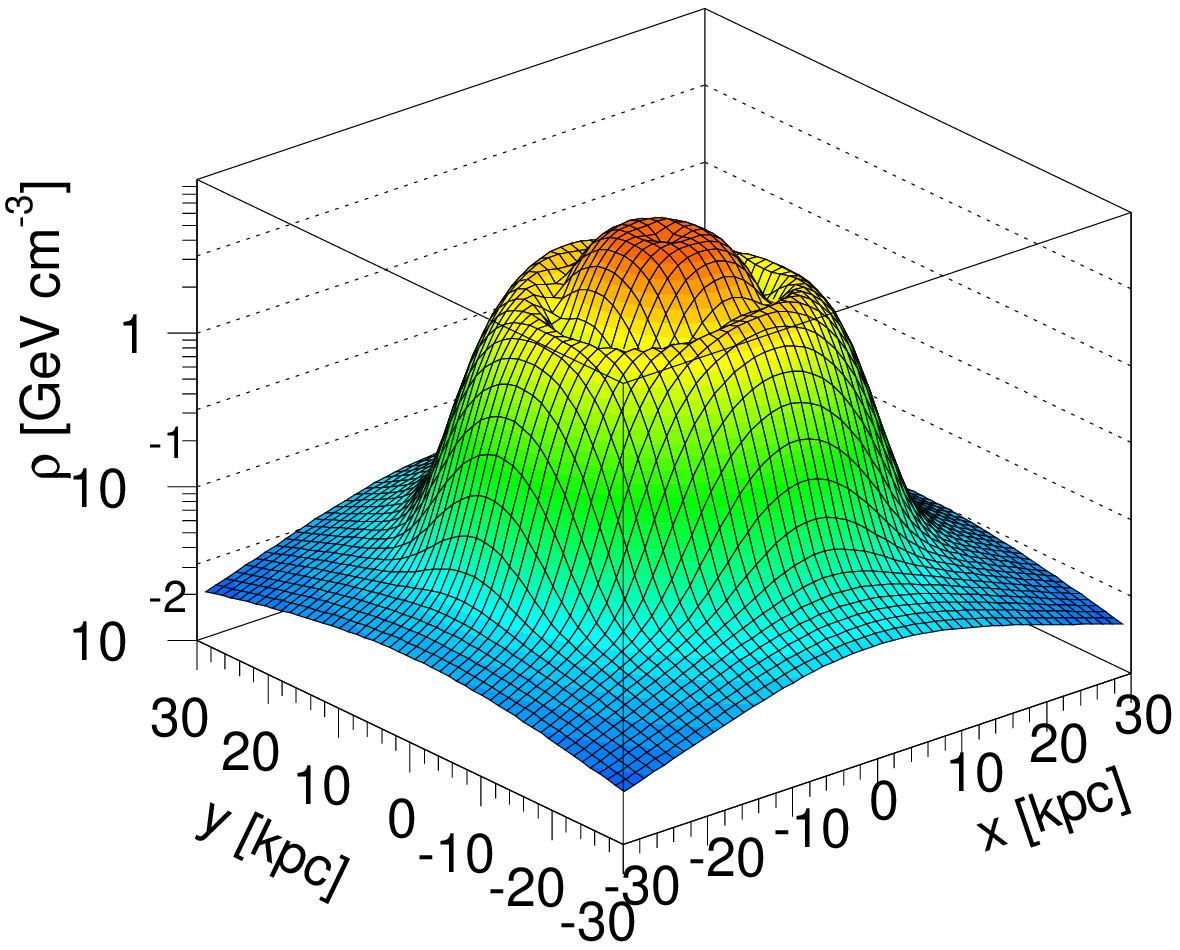}
\includegraphics[width=0.43\textwidth]{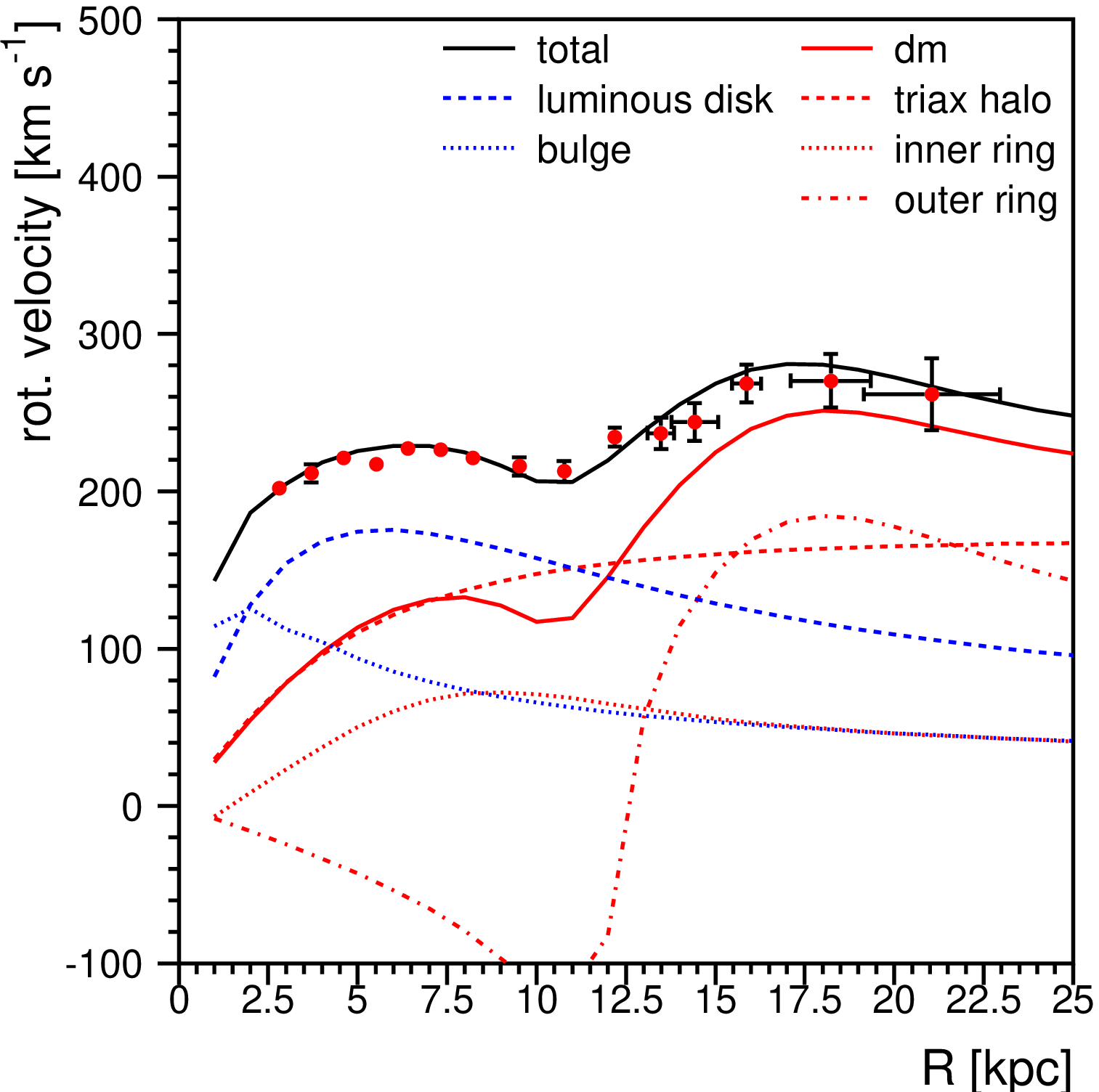}
\caption{CDM halo density in the galactic disk with the two ringlike
substructures at 4 and 14 kpc (left) and the corresponding rotation
curve (right). Adapted from Ref. \cite{deboer2}.} \label{f0}
\end{figure}
If the CR spectra are known and uniform in the Galaxy, the shape of
the background gamma rays is known from accelerator experiments. The
absolute value of neither the background nor DMA is known, because
of the large uncertainties in density of the interstellar medium, CR
density and CDM density. Therefore the obvious way to search for DMA
is to leave the absolute normalizations of the background and DMA
contributions free and fit only the  shapes of the background and
DMA for a given sky direction.

Experimentally, the spectral shape of the diffuse Galactic gamma
rays has been measured with the EGRET satellite;  we use the EGRET
data in the range 0.07 to 10 GeV in 8 energy bins.  For the relative
amount of electron- and nucleon-induced gamma rays the estimates
from real data, as implemented in the publicly available
``conventional'' GALPROP model\cite{galprop}, can be used, so one
has only one normalization constant for the background instead of
separate ones for the different background components.

Comparing the background with the EGRET data shows that above 1 GeV
there is a large deficit of gamma rays, which reaches more than a
factor of two towards the Galactic centre.\cite{hunter}  Fitting the
background together with the DMA, yields a perfect fit in {\it all}
sky directions for a CDM par\-ticle mass around 60
GeV.\cite{deboer1, deboer2} The shape fit automatically finds from
the free normalizations the relative amount of background and DMA.
Furthermore, the results are consistent with Supersymmetry
\cite{deboer3}.
From the amount of  excess in 180 independent sky directions one can
reconstruct the CDM profile, which in turn can be used to calculate
the rotation curve. The result explains   the hitherto unexplained
change of slope in the outer rotation curve\cite{deboer2}, as shown
in Fig. \ref{f0}. For the halo profile one is only interested in the
relative contributions in the various sky directions, so here all
experimental errors cancel, since the EGRET satellite does not care
in which direction it measures. The EGRET errors, as discussed in
Ref. \cite{Egret_errors} are not relevant, since we are not
interested in predicting absolute gamma ray fluxes, but only fit the
shapes with a free normalization. In this case only the
point-to-point errors are relevant. Furthermore, since the
systematic errors are dominating, every data point has approximately
the same weight, so changing the total error does not change the
solution for the minimum of the $\chi^2$ distribution; larger errors
only decrease its value. But in the fits of around 1400 data points
the $\chi^2/d.o.f$ is already well below 1 with a 7\% point-to-point
error, suggesting that these errors for a shape fit are already
overestimated.

\begin{figure}[t]
  \includegraphics[width=.47\textwidth]{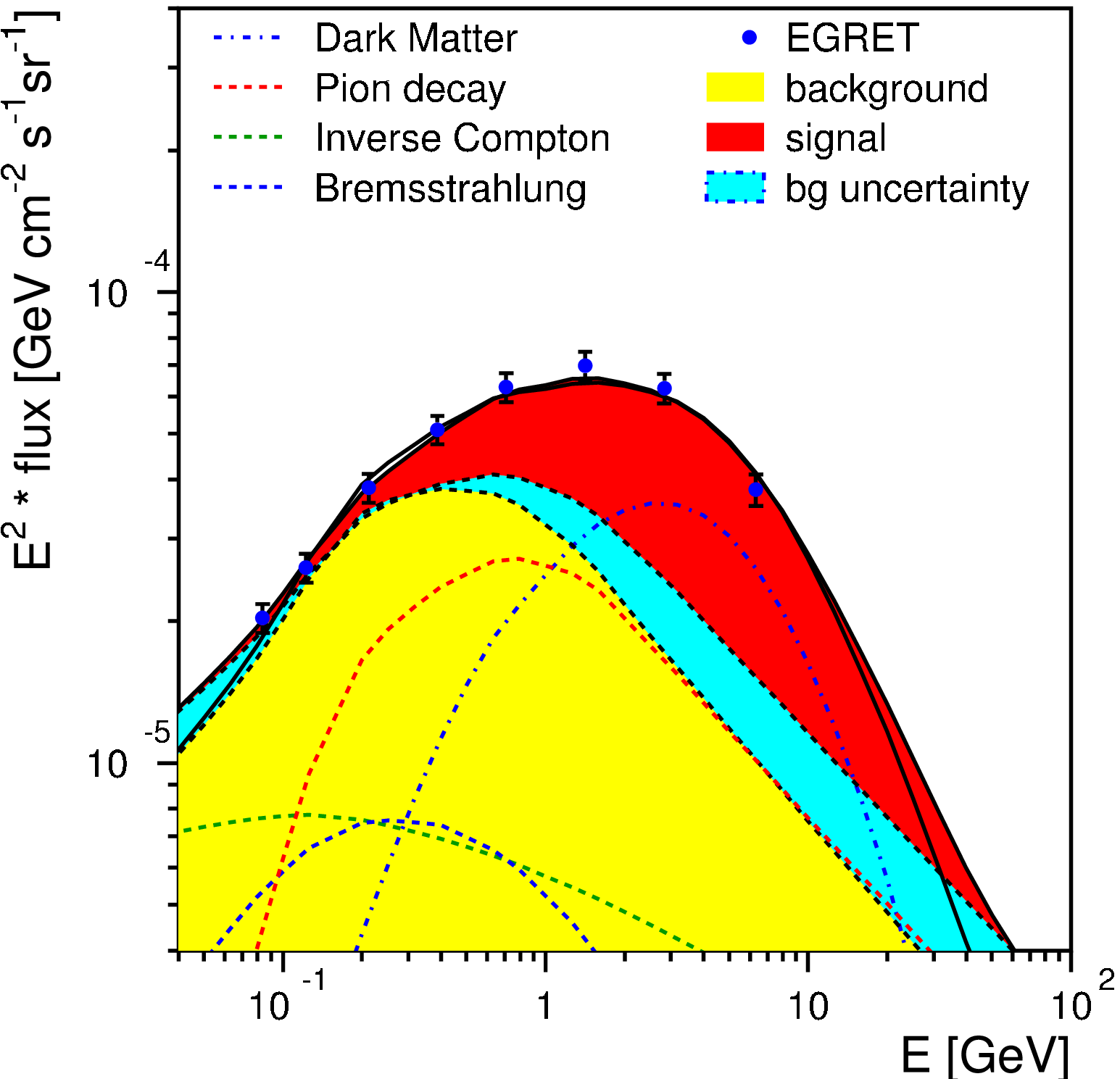}
  \includegraphics[height=.5\textwidth]{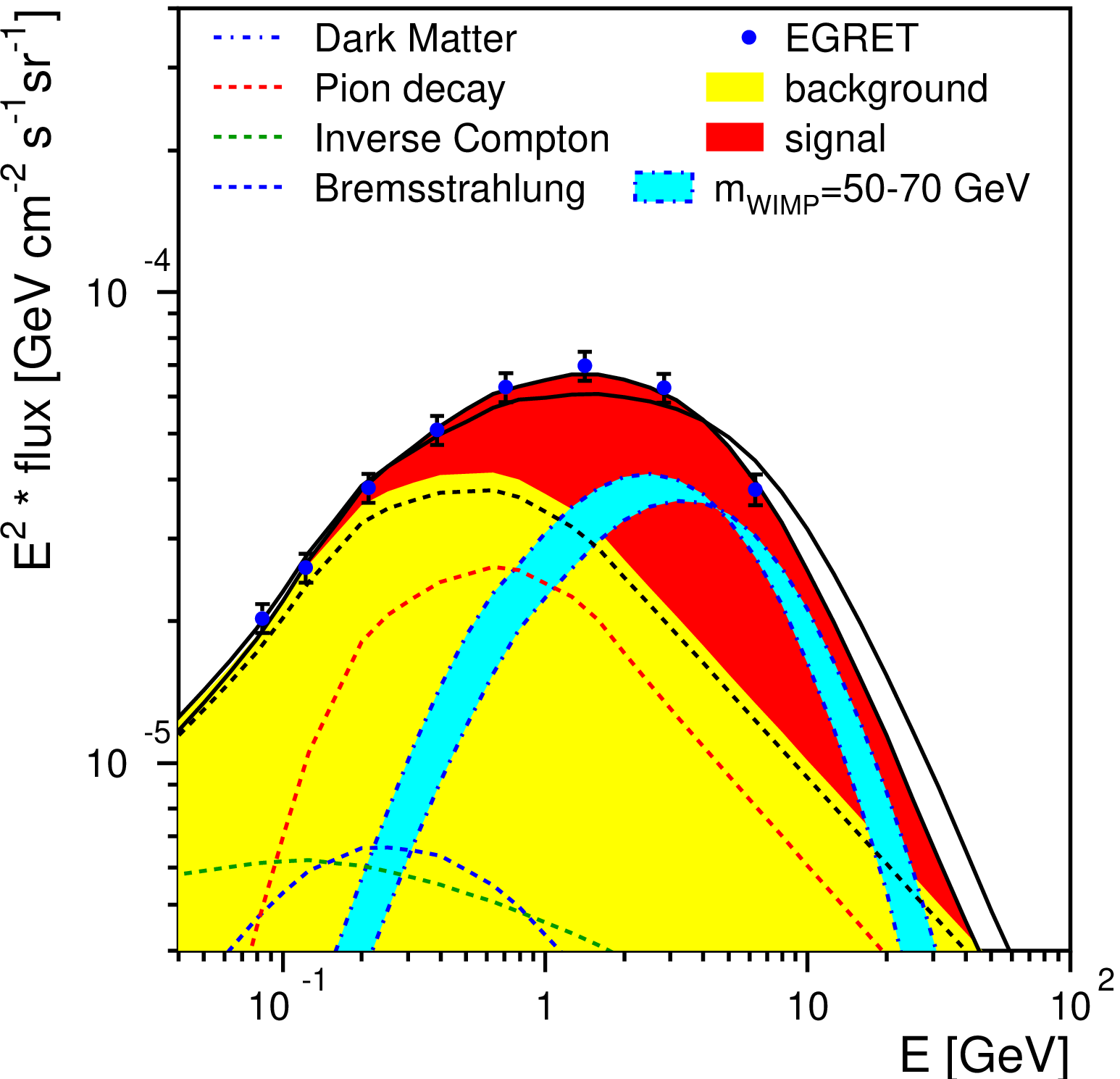}
  \caption{The EGRET gamma ray spectrum fitted with CDM annihilation for
  a 60 GeV WIMP mass (left).  The blue shaded area indicates the
  uncertainty from the shape of the CR spectra, which is dominated
  by the uncertainty in solar modulation (see text).
  On the right hand side the variation of the WIMP mass between 50 and 70 GeV
   is shown  (blue shaded area), which is the range allowed
  by the EGRET data assuming that the locally observed CR spectra are representative
  for our Galaxy.\label{f1}}
\end{figure}

Uncertainties from  the background, which are dominated by the solar
modulation uncertainty in the primary CR spectra,  are shown in Fig.
\ref{f1}. Note that the solar modulation depletes the CR spectrum at
low energies, but fitting the shape translates this into an
uncertainty mainly at high energy. This is simply because at low
energy the spectrum is almost purely background, so the expectations
are effectively ``normalized'' to this low energy data by the fit,
whatever the shape of the spectrum.

 Clearly the uncertainties in the background shape cannot explain
the excess, if one assumes that the locally observed CR spectra are
representative for the spectra outside the heliosphere after
correcting for  solar modulation. For nuclei the spectra are
expected to be indeed similar everywhere because the diffusion is
fast compared with energy loss times. So local variations of the
spectra or intensities, as proposed in Ref. \cite{optimized} to
explain the excess, seem to us unlikely, especially since this needs
in addition rather strong breaks in the CR injection spectra in
order to keep the gamma rays below 1 GeV the same, but only increase
the high energy gamma rays. Furthermore, these breaks are only
applied to protons, not to other nuclei in order to maintain the B/C
ratio. Also the ``fresh'' harder source component $\propto E^{-2}$
instead of $\propto E^{-2.7}$ is not expected to yield a significant
effect, since this is only a small fraction of the total CR density.
This is obvious for older Galaxies, where the amount of CR escaping
to outer space (with an escape time of $\cal{O}$$(10^7-10^8)$ y)
should be equal to the amount of generated CR (with a source life
time of $\cal{O}$$(10^4-10^5)$ y), so the fresh component should be
of $\cal{O}$$(10^{-2}-10^{-3})$. That the shape of CR in the steady
state is similar everywhere, is confirmed by a numerical solution of
the diffusion equation, as used in GALPROP.  For this
``conventional'' model of CRs having everywhere the spectrum of the
locally observed one, the WIMP mass is rather well constrained
(50-70 GeV), as shown on the right hand side of Fig. \ref{f1}.


In summary, the gamma rays play  a very special role for indirect
CDM searches, since they point back to the source and are
independent of propagation models. Therefore the gamma rays provide
a perfect means to reconstruct the intensity (halo) profile of the
CDM by observing the intensity of the gamma ray emissions in the
various sky directions. This halo profile can in turn be used to
check the shape of the rotation curve, thus providing a direct link
between the excess of the gamma rays and the strongest evidence for
CDM, the rotation curve.

\section{Antiproton fluxes}

Contrary to gamma rays the charged particles change their direction
by the interstellar magnetic fields, energy losses and scattering.
Therefore one needs a detailed propagation model to calculate the
amount of particles which will arrive from the source to the
detector. Charged particles usually make a random walk process by
changing their direction through interaction with the galactic
magnetic field, which is thought to have a larger  random
(turbulent) component in the interstellar space. But galactic winds
may lead to a strong convective transport of these magnetic
turbulences perpendicular to either side of the galactic plane,
which take the charged particles with them to outer
space\cite{ptuskin}, thus leading to strong anisotropic propagation.
Furthermore, the transport and production of charged particles can
be strongly influenced by the neighbourhood of our solar system with
its underdense local bubble and overdense clouds (``local fluff'')
and magnetic walls with as much as five orders of magnitude drop in
the diffusion coefficient in the heliosphere.\cite{florinski}   Up
to now all these highly uncertain details have not been studied. The
common propagation models simply assume an isotropic diffusion in
all directions in a large volume before the particles escape to
outer space.
\begin{figure}[t]
\includegraphics[width=0.46\textwidth]{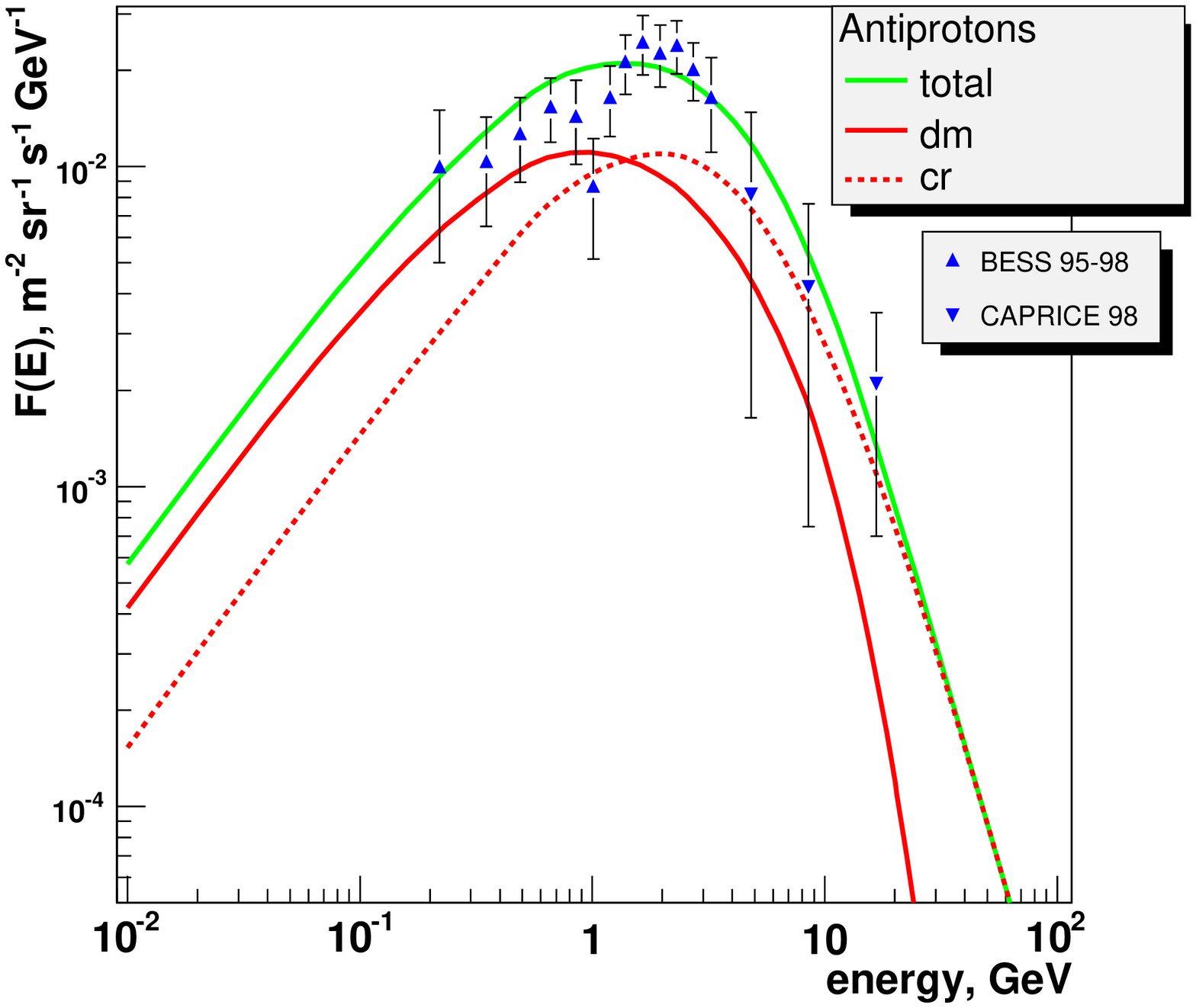}
\includegraphics[width=0.46\textwidth]{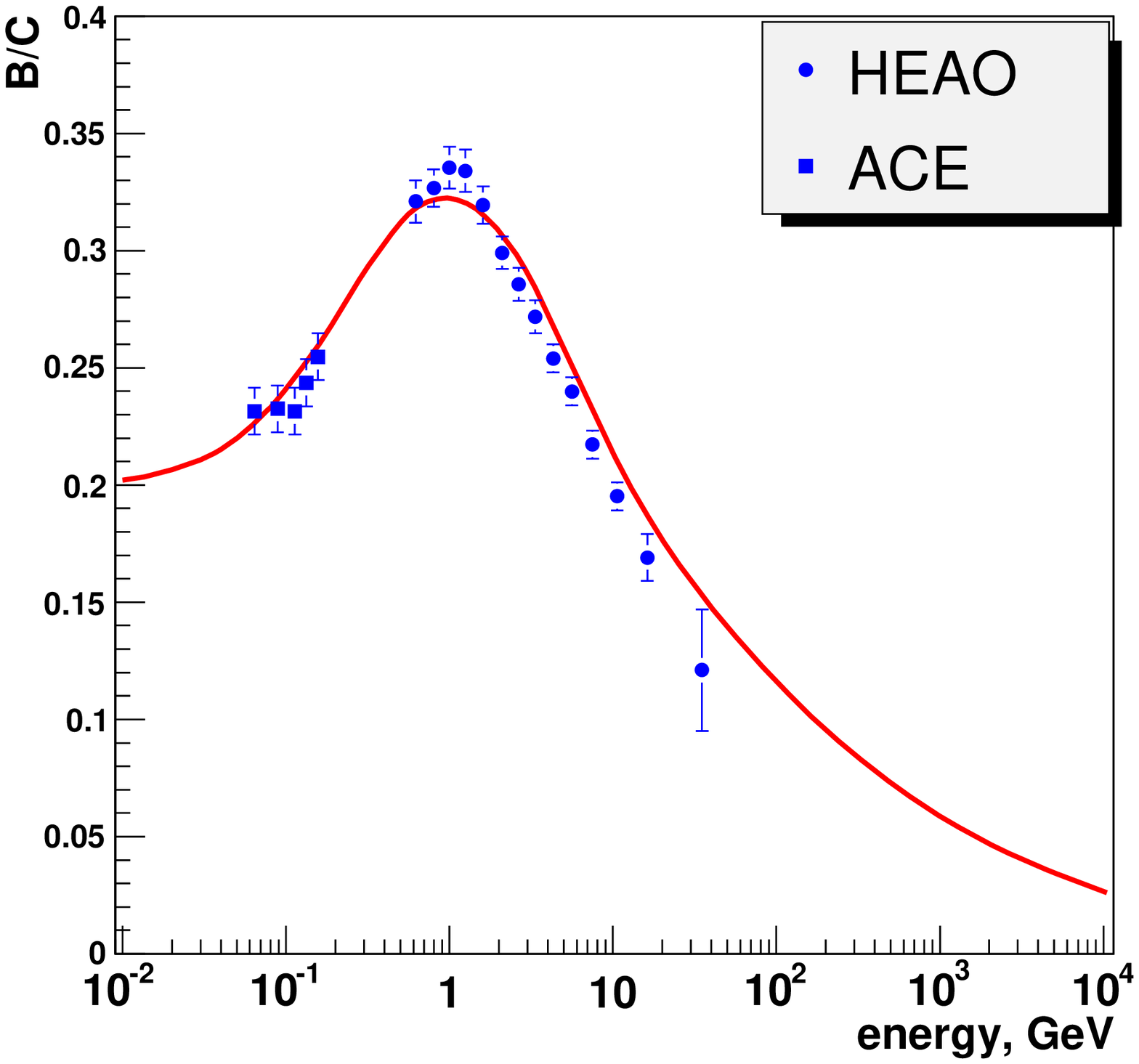}
\caption{The antiproton fluxes and the B/C ratio from the modified
GALPROP code including DMA and anisotropic propagation. Note that
roughly half of the antiprotons are coming from DMA, as for the
gamma rays above 1 GeV, while DMA does not contribute to the B/C
ratio.} \label{f2}
\end{figure}

By increasing the convection perpendicular to the disk and
implementing the local bubble,  the local clouds and ``magnetic
walls'' with slow diffusion in the solar neighbourhood one can
change the antiproton flux by an order of magnitude, while still
being consistent with the B/C ratio, as shown in Fig. \ref{f2}.  The
DMA contribution explains the traditional EGRET "excess" of gamma
rays without the need for assuming that the locally observed CR
spectra are different from the CRs in the rest of the Galaxy.  Here
the GALPROP model was used after including DMA and retuning the
diffusion and convection parameters. Traditionally these parameters
have been determined by the B/C ratio and the cosmic clocks, like
the $^{10}Be/^9Be$ ratio. The diffusion coefficient needed for the
B/C ratio required a large halo with a distance of z=4 kpc to get a
long enough trapping time for the cosmic clocks. This traps also the
antiprotons from DMA, thus leading to the solution from Ref.
\cite{bergstrom} using DarkSusy. This results can be reproduced with
GALPROP, if the isotropic diffusion dominates. In our case the
antiprotons are blown to outer space by convection, which overtakes
diffusion a few hundred parsec above the disk. In this case of large
convection a one to two orders of magnitude smaller diffusion
coefficient is needed, which is much closer to the values used in
heliospheric propagation models.\cite{florinski} The large B/C ratio
is obtained by the local fluff with a size of 5 pc in the local
bubble. Note that most of the molecular gas is concentrated in large
molecular clouds, which occupy only a few \% of the volume. These
clouds act as localized sources  of all secondary particles and are
particularly strong if nearby, since the flux decreases as $1/r^2$.
Naively one expects that if locally a large amount of secondary
boron nuclei  are produced (by fragmentation of CNO and heavier
atoms on the gas), one expects a correspondingly large amount of
secondary antiprotons. This is not true, since the latter require CR
protons with an energy above 10 GeV (due to threshold effects),
while for rigidities of CNO nuclei below 10 GeV the fragmentation
cross sections just increase. Therefore changing the injection
spectra of primary particles below 10 GeV by 10\% or changing the
energy dependence of the diffusion constants immediately changes the
antiproton/B ratio for rigidities around 1 GeV by a factor of a few.

 In summary, recent claims that the antiproton fluxes
exclude the DMA interpretation of the EGRET excess should be
considered in the light of the limitations of DarkSusy, which uses a
simple analytical solution of the diffusion equation with
unrealistic smooth gas distributions and isotropic diffusion
coefficients. In order to allow for anisotropies in gas
distributions, convection velocities and diffusion coefficients one
has to resort to numerical solutions of the diffusion equation, as
implemented in GALPROP after suitable modifications for DMA,
non-equidistant grids and anisotropic diffusion and convection, i.e.
D(r,z) and V(r,z). In the latter case a consistent flux of local CR
spectra, antiproton fluxes, B/C ratio and gamma rays can be
obtained.

\section{Summary and Outlook}
In summary, the excess of EGRET diffuse gamma rays has all the
properties expected for DMA. Especially the excess has the shape
expected for the annihilation of  60 GeV WIMPs and the distribution
of the excess over the sky is in perfect agreement with the shape of
the rotation curve of our Galaxy, which is the hallmark of a DMA
signal.

  Objections
against the DMA interpretation of the EGRET excess
\cite{bergstrom_pb} concerning a too high antiproton flux should be
considered in the light of their simple diffusion model. Our
preliminary investigations show that  a more realistic  propagation
model can reduce the antiproton flux by more than an order of
magnitude. Therefore it is hard to use antiprotons to  search for
{\it light} CDM particles, which yield a soft antiproton spectrum
similar to the background. However, the antiprotons are  perfect to
tune the many parameters in more realistic pro\-pagation models, if
the CDM halo is determined from the gamma rays.

Future data on high energy gamma rays (GLAST satellite) and high
energy charged particles (space experiments PAMELA and AMS) will be
of great interest in order to see if  this picture of DMA is
confirmed, while direct DM detection experiments and the new hadron
collider LHC may be able to determine independently the WIMP mass.
If they all find a WIMP mass in the range suggested by the EGRET
excess, this would be great.
\section{Acknowledgements}
 Very helpful discussions with V. Florinski, V.Ptuskin,
 R. Schlickeiser and H. V\"olk about propagation
uncertainties are acknowledged. This work was supported by the DLR
(Deutsches Zentrum f\"ur Luft- und Raumfahrt).

\end{document}